\documentclass{article}
\usepackage{setspace}
\usepackage[english]{babel}
\usepackage[utf8]{inputenc}
\usepackage{amsmath}
\usepackage{amsfonts}
\usepackage{amssymb}
\usepackage{amsthm}
\usepackage{cancel}
\usepackage{slashed}
\usepackage{graphics}
\usepackage{graphicx}
\usepackage{fancyhdr}
\usepackage{multirow}
\usepackage{float}
\usepackage{breqn}
\usepackage{pdfpages}

\usepackage{hyperref}
\usepackage{fancyhdr}
\usepackage{bm}

\usepackage[left=3cm,top=2.25cm,bottom=2.75cm,right=2.5cm,includehead]{geometry}
\usepackage[titletoc]{appendix}
\usepackage{youngtab}
\usepackage{dsfont}
\usepackage{tikz}
\usetikzlibrary{shapes,arrows}
\usepackage{bbm}
\usepackage{mcite}
\usepackage{cite}
\usepackage{verbatim}
\usepackage{indentfirst}

\hypersetup{
    colorlinks=true,
    linkcolor=blue,
    urlcolor=cyan,
    citecolor=cyan,}

\title{Anomalous transport from geometry}
\author{Karl Landsteiner${}^1$, Sergio Morales-Tejera${}^{1,2}$ and Pablo Saura-Bastida${}^{1,3}$}

\begin{document}

\maketitle\thispagestyle{empty}
{\begin{center}${}^1$Instituto de F\'isica Te\'orica UAM/CSIC, c/Nicol\'as Cabrera 13-15, Universidad Aut\'onoma de Madrid, Cantoblanco, 28049 Madrid, Spain\\
${}^2$Departamento de F\'isica Te\'orica, Universidad Aut{\'o}noma de Madrid, Campus de Cantoblanco, 28049 Madrid, Spain\\
${}^3$ Departamento de Autom\'atica, Ingenier\'ia El\'ectrica y Tecnolog\'ia Electr\'onica, Universidad Polit\'ecnica de Cartagena, Calle Dr. Fleming, S/N, 30202 Cartagena, Murcia 
\end{center}}

\begin{abstract}
  We revisit the relation between black hole geometries and chiral transport.
  Integrating the anomaly equation in a black hole geometry allows to derive the chiral transport coefficients for the thermal gas far from the horizon.
  The key ingredient is to impose vanishing of the covariant current on the horizon. We extend the method to include the usual gauge anomaly for charged black holes and to weak magnetic fields. This allows to derive the full set of transport coefficients describing the chiral magnetic and chiral vortical effects.
\end{abstract}
\vfill
{\tt
\href{mailto:karl.landsteiner@csic.es}{karl.landsteiner@csic.es}\\[-5pt] 
\href{mailto:pablo.saura@upct.es}{pablo.saura@upct.es}\\[-5pt]
\href{mailto:sergio.moralest@uam.es}{sergio.moralest@uam.es} 
}
{\hfill IFT-UAM/CSIC-22-155 }

\newpage
\tableofcontents

\newpage

\section{Introduction}
Chiral anomalies are among the most emblematic features of the quantum field theory of relativistic fermions \cite{Bertlmann:1996xk}. Gauge symmetries have
to be free of anomalies and this leads to severe constraints on the fermion content of quantum gauge theories.
A somewhat different aspect of chiral anomalies is that they also give rise to dissipationless currents when theories
containing chiral fermions are put into a finite temperature and finite density (near-)equilibrium state (see \cite{Landsteiner:2016led}, \cite{Kharzeev:2013ffa} for reviews).

A subtle issue of chiral anomalies is that they come in consistent and covariant forms \cite{Bardeen:1984pm}. Consistent anomalies are
defined via the Wess-Zumino consistency condition, meaning that the anomaly is integrable and stems from a (non-local) 
functional, the 1PI-effective action $\Gamma_{1PI}[A,g]$ in which the gauge field $A_\mu$ and the metric $g_{\mu\nu}$ serve as sources for
the current $J^\mu$ and the energy momentum tensor $T^{\mu\nu}$. Typically the precise form of consistent anomalies depends
on a choice of local counterterms that can be added to the 1PI-effective action. A different way of defining anomalies
is via their covariant form. There, one demands that the quantum operators $J^\mu$ and $T^{\mu\nu}$ are themselves 
covariant objects under the symmetry transformations (even the anomalous ones). The covariant anomalies in the conservation equations for the currents and energy-momentum tensor are

\begin{align}
\label{eq:anomalyJ}
& \nabla_\mu J^\mu_{a} =  -\frac{d_{abc}}{32 \pi^2} \epsilon^{\mu\nu\sigma\rho} F^b_{\mu\nu}F^c_{\sigma \rho} - \frac{b_a}{768 \pi^2} \epsilon^{\mu\nu\sigma\rho} R^\alpha_{\:\:\beta\mu\nu}R^\beta_{\:\:\alpha\sigma \rho}\quad , \\
\label{eq:anomalyT}
& \nabla_\mu T^{\mu \nu} = F^{a,\nu \mu} J_{a,\mu} - \frac{b_a}{384 \pi^2} \epsilon^{\sigma \rho \alpha \beta} \nabla_\mu \left(F^a_{\sigma \rho} R^{\nu \mu}_{\quad \alpha \beta}\right) \quad .
\end{align}

We should stress here that the gauge fields $A^a_\mu$ as well as the metric $g_{\mu\nu}$ are viewed as classical fields in 
these equations. The presence of the anomalies is completely determined by the anomaly coefficients $d_{abc} = \sum_{r} q_a^r q^r_b q^r_c - \sum_{l} q_a^l q^l_b q^l_c$ and $b_a = \sum_{r} q_a^r - \sum_{l} q_a^l $, where $q^{r,l}_a$ is the charge of the right- (left-) handed fermion under the symmetry indexed by $a$\footnote{We restrict here to the relevant case of having $U(1)$ symmetries only.}.
 
It has been noticed long time ago that chiral fermions at finite density and chemical potential subject to a magnetic field or to rotation give rise to non-vanishing expectation values \cite{Vilenkin:1980fu, Vilenkin:1979ui}. Furthermore it turns out that these currents are
completely determined by the anomaly coefficients $d_{abc}$ and $b_a$.
\begin{align}
\label{eq:cmecveJ}
& \vec j_a =  \frac{d_{abc} \mu_b}{4\pi^2} \vec B_c + \left(\frac{d_{abc}\mu_b\mu_c}{4\pi^2} + \frac{b_a T^2}{12}\right) \vec \Omega \,, \\
\label{eq:cmecveT}
& \vec j_\epsilon =  \left(\frac{d_{abc}\mu_b\mu_c}{8\pi^2} + \frac{b_a T^2}{24}\right)\vec B_a +  \left(\frac{d_{abc}\mu_a\mu_b\mu_c}{8\pi^2} + \frac{b_a \mu_a T^2}{24}\right)\vec \Omega\quad ,
\end{align}
where $\vec B_a = \vec\nabla \times \vec A_a$ is a magnetic field of the gauge field coupling to the current $J^\mu_a$ and $\vec\Omega$ is the vector of angular velocity. These are known as chiral magnetic and chiral vortical effects respectively.

It has been realised rather early that the terms proportional to only the chemical potentials are determined by the anomaly \cite{Giovannini:1997eg, Alekseev:1998ds, Fukushima:2008xe, Son:2009tf, Neiman:2010zi}. 
A typical argument put forward first in \cite{Son:2009tf} relies on constructing
an entropy current and demanding that the entropy production rate is positive definite. The temperature dependent parts enter this construction only through undetermined integration constants \cite{Neiman:2010zi}.
The reason for this apparent difference is that the hydrodynamic framework relies on a derivative counting scheme in which the gravitational terms are higher order terms \footnote{
The naive derivative counting argument suggests that the gravitational anomaly gives rise to transport at third order. Such transport coefficients have indeed be discussed recently in \cite{Prokhorov:2022udo, Prokhorov:2022snx} and earlier in \cite{Kimura:2011ef}.} and thus
do not appear in equations (\ref{eq:cmecveJ}), (\ref{eq:cmecveT}) which are both
of first order in the derivative counting scheme (since $\vec B_a =\vec \nabla \times \vec A_a$ and $\vec \Omega = \vec \nabla \times \vec v$ with $\vec v$ the velocity field).

Explicit calculations in field theory and in strongly coupled models 
based on the gauge-duality do however indicate that the temperature dependent terms are induced by the gravitational parts of the chiral anomaly \cite{Landsteiner:2011cp,Landsteiner:2011iq}. 
In particular, the holographic calculation including the gravitational anomaly is not different from previous ones including only the gauge field anomalies \cite{Erdmenger:2008rm, Banerjee:2008th}.
This should be taken as a strong hint that despite the apparent mismatch in derivatives the terms with chemical potential and with temperature follow from the same principles.

It has first been shown that in Euclidean space there is a way of relating
the gravitational anomaly terms to the induced current by considering the quantum field theory on a non-trivial geometry that interpolated between empty flat space and finite temperature viewed as the field theory living on
$\mathbb{R}^3\times S^1$, where the length of the circle is interpreted as 
the inverse temperature \cite{Jensen:2012kj}. Later a Lorentzian construction was put forward in \cite{Stone:2018zel}. In Lorentzian terms the geometries that interpolate between
flat space and finite temperature can be interpreted as black hole geometries. Indeed, a free falling observer will note nothing special upon
crossing the event horizon of a black hole. On the other hand, it is known that a static asymptotic observer at a large distance from the black hole will observe a constant flux of thermal Hawking radiation. Thus, a quantum field theory place in such a black hole background will be in equilibrium
with thermal radiation at infinity but experience no temperature at the 
horizon. It was shown in \cite{Stone:2018zel} that by integrating the gravitational parts of the anomaly equations between the horizon and infinity the temperature dependent parts of the currents can be obtained. 

This derivation is very different from the entropy considerations based on
hydrodynamics. One is then left with the impression that the two contributions to the chiral anomaly are on very different footing. The purpose of this letter
is to close a gap in the literature and to
show that the argument based on black hole geometries is in fact
more general and also allows to recover all the the anomaly induced currents that
depend on the chemical potential. Thus from this point of view there is really no difference in the two types of anomalies.  

The paper is structured as follows. In section two we first review the construction of \cite{Stone:2018zel}. This is basically a two-dimensional argument in which the background magnetic field is imagined to be infinitely strong such that the effective dynamics is completely determined by the fermions in the lowest Landau level. In contrast the chiral vortical effect is derived to first order in the vorticity $\vec \Omega$.
In section 3 we consider electrically and magnetically charged black holes in $3+1$ dimensions. We show that the temperature and chemical potential dependent terms in the currents can be obtained at once by integrating the anomaly equations. In section 4 we show that it is also possible to do this for an electrically charged and rotating black hole, thus allowing us to  recover the chiral vortical effect in the energy current. 
We close in section 5 with some conclusions and outlook.

\newpage
\section{Integrating the anomaly on a black hole geometry}

From the point of view of quantum field theory the anomalous conservation equations (\ref{eq:anomalyJ}), (\ref{eq:anomalyT}) are operator relations which hold for arbitrary expectation values of matrix elements.
On the other hand the transport phenomena described by (\ref{eq:cmecveJ}) and (\ref{eq:cmecveT}) are tied to a particular state in the theory. This is manifest in the explicit expressions for the CME and CVE, which are written in terms of state dependent quantities: the temperature $T$ and the chemical potentials $\mu_a$. To establish a precise connection between both perspectives one needs to engineer a background field configuration that mimics appropriately the state, and then integrate the anomalous conservation equation in this background\footnote{We would like to point out that the anomalous currents are produced by quantized fields on top of the given background, not by the matter generating the background gravitational field.}. This is precisely what a generic black hole does. The black hole is generic in the sense that it needs not satisfy any equations of motion. Only
the generic (topological) properties of the presence of a non-degenerate horizon and an
asymptotically flat region connected smoothly are important. As we will see
a chemical potential may be introduced via a "generic" gauge field by considering charged
black holes. 

\par

We briefly review now the anomaly integration method put forward in \cite{Stone:2018zel}, a similar method was originally presented in \cite{Robinson:2005pd}.

\subsection{Thermal CME in 1+1 dimensions}
\label{sec:2.1}

First one considers a system in a strong magnetic field. Fermions will then
populate the lowest Landau level and therefore the relevant physics will be captured by an effective $1+1$ dimensional model. In two spacetime dimensions the mixed four dimensional anomaly for the energy momentum tensor reduces to a pure gravitational anomaly given by: 

\begin{equation} \label{2-Anom-2D}
\nabla_\mu T^{\mu \nu}_{R,\:L} = \mp c_g \epsilon^{\nu \sigma} \partial_\sigma R \,, 
\end{equation}

\noindent
where $R$ is the Ricci scalar, $\epsilon^{\nu\sigma}=\epsilon(\nu\sigma)/\sqrt{-g}$ is the two dimensional Levi-Civita tensor and $\epsilon(\nu\sigma)$ the totally antisymmetric symbol. 
For reasons of generality we also allow for a completely general anomaly 
coefficient $c_g$. For two dimensional chiral conformal field theories $c_g$
can be identified with the  central charge.
The subscripts $R,L$ denote right- or left-handed particles respectively, and the $\pm$ is correlated with the subscript. 
The central idea of \cite{Jensen:2012kj,Stone:2018zel} is to integrate the anomaly equation (\ref{2-Anom-2D}) on 
a background that interpolates between the vacuum and the thermal state.
With Lorentzian signature such a background is given by a $1+1$ dimensional black hole with Hawking temperature $T$. 
The theory at large distance from the horizon must be in equilibrium with the Hawking radiation emitted by the black hole, thus the current will be that of a theory at finite temperature with $T$ the Hawking temperature (see also \cite{Robinson:2005pd}). A generic metric for a two dimensional black hole is:

\begin{equation}
ds^2 = -f(r) dt^2 + \frac{1}{f(r)} dr^2 \quad .
\end{equation}

We are considering this metric as a non-dynamical metric, so it needs not satisfy Einstein's equations. However, we do demand that there is a horizon $f(r_h) = 0$ for some $r_h$ and that the metric is asymptotically flat at infinity $f(r \to \infty) = 1 $. The Ricci scalar is given $R=-f''(r)$, where prime denotes derivative with respect to $r$, so asymptotic flatness further requires $f''(r \to \infty)=0\,$. We may also demand that this function goes smoothly to $1$ at infinity, that is $f^\prime(r \to \infty) = 0$. With this metric, we may define the energy current as:

\begin{equation}
\label{eq:Energycurrent}
J^\mu_E = T^{\mu \nu}_{R,\:L} \eta_\nu \,,
\end{equation}

\noindent
where $\eta = \partial_t$ is a Killing vector field of the chosen metric. 
The anomaly equation (\ref{2-Anom-2D}) becomes for the energy current:

\begin{equation}
\nabla_\mu J^\mu_E = \mp c_g \eta_\nu \epsilon^{\nu \sigma}
\partial_\sigma R \quad .
\end{equation}
 
We can explicitly compute the right hand side of the last equation in the
black hole background and notice that the term can be written in terms of a total derivative: 
\begin{equation}
\nabla_\mu J^\mu_E =  c_g \partial_r \left[ f(r)   f^{\prime \prime} (r) - \frac{1}{2}  f^{\prime 2} (r) \right]  \, .
\end{equation} 
Thus we obtain an expression for the energy current that we expect to reproduce the CME. Only radial dependence is assumed. Consequently we solve for the radial component of the current. As the r.h.s. of the equation is a total derivative, the final result is going to depend just on the boundary conditions we have imposed for the function $f(r)$. This is of course expected due to the 
topological nature of anomalies.

\par 

We can then arrive to the current of the chiral magnetic effect by integrating the anomaly equation in this background. One should demand that the covariant form of the current vanish at the horizon \cite{Stone:2018zel}\footnote{This
is equivalent to the requirement of "consistency with the Euclidean vacuum"
introduced in \cite{Jensen:2012kj,Jensen:2013rga}.}. Equivalently we may argue that the norm of the (covariant)  current appearing in (\ref{eq:Energycurrent}) should not blow up in any regular point of the spacetime. In particular $J_{\mu}J^{\mu} \propto \frac{(J^r)^2}{f(r)}$. This is regular only if the current appearing in (\ref{eq:Energycurrent}) vanishes at the horizon. On the other hand, it is at infinity that one observes the thermal radiation, and hence one expects to reproduce the correct result at infinity. Similar arguments apply in subsequent sections. Assuming time independence one finds for the current at infinity

\begin{equation}
J_{E}^r|_{r=\infty} =  8 c_g \pi^2 T^2 \, ,
\end{equation}

\noindent
where $T$ is the Hawking temperature defined by the surface gravity via:
\begin{equation}
    T = \frac{\kappa}{2\pi} = \frac{f^\prime(r_h)}{4\pi} \,.
\end{equation}

The magnetic field is absent as a consequence of the dimensional reduction. It may be reinstated by noting that the 
density of states in the lowest Landau level is given by $\frac{B}{2\pi}$.
If we also specialize to the case of a single chiral fermion such that
$c_g = \pm\frac{1}{96\pi}$ depending on chirality, we get 

\begin{equation}\label{eq:cmeTB}
\vec{j}_{E;\:R,\:L} = \pm \frac{T^2}{24}\vec{B} \, .
\end{equation}
This is indeed the known result for the chiral magnetic effect of a chiral fermion at finite temperature in equation (\ref{eq:cmecveT}).
Let us comment now to a subtle issue related to the chiral magnetic effect. It is often
argued that the CME can be understood easily as a consequence of the physics of the Lowest Landau level. Fermions in the Lowest Landau level do behave as effective $1+1$ dimensional
particles and thus the CME can be viewed as a consequence of the operator relation
$J^\mu_5 = \epsilon^{\mu\nu} J_\mu$ valid in two spacetime dimensions\footnote{This can also be written as $J^\mu_{R,L} = \pm \epsilon^{\mu\nu} J_{R,L,\mu}$ in terms of right- and left-handed currents.}. This operator relation
basically states that the current is given by the axial charge and vice versa.
The reduction to $1+1$ dimensions however relies on the presence of a strong magnetic field. If this reduction would be essential for the CME then one should expect corrections
to the the CME formula for high temperature and small magnetic field. It is however well known that the CME
is an exact formula in $3+1$ dimensions and no such corrections arise. Therefore it is necessary to generalize this the argument leading to (\ref{eq:cmeTB}) to four dimensional black holes and small magnetic fields.

\subsection{Thermal CVE in 3+1 dimensions}
\label{sec:2.2}

This strategy can also be adopted to compute the thermal part of the CVE in $3+1$ dimensions \footnote{In this case the dimensional reduction would amount to including a $U(1)$ symmetry associated to the axial symmetry of the system \cite{Murata:2006pt}}. Inspired by the Kerr metric, the authors of \cite{Stone:2018zel} propose to work with the following background metric:

\begin{equation}
\label{eq:CVEbackground}
ds^2 = -f(z) \frac{(dt - \Omega \rho^2 d\phi)^2}{1-\Omega^2\rho^2} + \frac{dz^2}{f(z)} + d\rho^2 + \frac{ \rho^2(d\phi - \Omega dt)^2}{1-\Omega^2\rho^2} \,,
\end{equation} 

\noindent
which is in cylindrical coordinates and constructed such that it asymptotes to  Minkowski space. Notice that the background encodes rotation, where $\Omega$ is the constant angular velocity of the event horizon. Again one imposes the existence of a horizon ($f(z_h)=0$ for some $z_h$) and asymptotic flatness: $f(z\to \infty)=1\,$ and $f'(z\to \infty)=f''(z\to \infty)=0\,$. 

Contrary to the previous case, here the first non-trivial contribution arises in the $U(1)$ current $J^{\mu}$ whose anomalous conservation equation in the presence of gravity is proportional to the Pontryagin density:

\begin{equation}
    \nabla_{\mu}J^{\mu} = -c_g \epsilon^{\mu\nu\rho\sigma}R^{\alpha}_{\,\,\beta\mu\nu}R^{\beta}_{\,\,\alpha\rho\sigma}
\end{equation}

It is not hard to show that this equation admits a static solution with $z$-dependence only in which only the $z$-th component of the current is non-vanishing. 
We also keep only the first order in an expansion in the angular velocity
$\Omega$ since there is no indication that higher order terms are universal.
One finds

\begin{equation}
\nabla_\mu J^\mu = - 8 c_g f^\prime(z) f^{\prime\prime}(z)
\Omega + \mathcal{O}(\Omega^3)\, .
\end{equation}

\noindent
Integrating it and imposing vanishing of the current at the horizon yields
to lowest order in $\Omega$ \footnote{It is worth mentioning that we obtain a result depending on rotation at infinity although our metric is asymptotically Minkowski, and thus does not rotate at infinity. This is an effect of the coordinates we have chosen, in this specific choice the rotation is at the horizon, but one can perform a change of coordinates to relocate it at infinity, in that case the result would still remain the same because physics cannot depend on the coordinates one have chosen. In the coordinates in which rotation is present at infinity the chiral vortical effect is better described as a chiral gravito-magnetic effect \cite{Landsteiner:2011tg}.}

\begin{equation}
J^z (z\to \infty) = 4 f^{\prime 2}(z_h) \Omega =  64 \pi^2 c_g  T^2\Omega \, ,
\end{equation}

\noindent
which for a single chiral fermion gives $\vec j = \pm \frac{T^2}{12} \vec \Omega$. This coincides with the expression in equation (\ref{eq:cmecveJ}) specified to a single chiral fermion.

\section{Chiral Magnetic Effect in 3+1 dimensions via Anomaly}\label{sec:cme}

In this section, we are going to present a detailed derivation of the CME for a $U(1)$  current and the energy current using the anomaly integration method explained in the previous section. We consider black hole geometries that interpolate between the vacuum with vanishing currents at the horizon and the state with both chemical potential $\mu$ (of the given $U(1)$ symmetry) and temperature $T$ in presence of a magnetic field $B\,$. Thus we consider electrically and magnetically charged black hole geometries \cite{Hawking:1995ap} hints that a suitable background is:

\begin{align}
\begin{split}
\label{eq:CMEbackground}
& ds^2 =  -f(r) dt^2 + \frac{1}{f(r)}dr^2 + r^2 d\Omega_{(2)}^2 \quad ,
\\
& A = A_t(r)dt - Q \cos \theta d\phi\,,
\end{split}
\end{align}

\noindent
where $Q$ is the magnetic charge and we are working in spherical coordinates. We fix part of the gauge freedom by using the radial gauge $A_r=0\,$. The magnetic field is radial and arises from the $\phi$ component of the gauge field. By definition the chemical potential is the energy required to thermalize a unit of charge. In the 
geometry (\ref{eq:CMEbackground}) this is the energy needed to bring a unit of charge from infinity to behind the horizon
\begin{equation}
\label{eq:chemical}
    \mu = \int_{r_H}^
\infty F_{\nu\mu}\eta^{\mu}dx^{\nu}\,,
\end{equation}

\noindent
where $\eta^{\mu}$ is a timelike killing vector and $F_{\mu\nu}$ is the field strength of the gauge field: $F=dA\,$. Using the Killing vector of the background metric $\eta=\partial_t$ we find $\mu=A_t(\infty)-A_t(r_h)\,$. As we have restricted ourselves to the radial gauge and only want to consider radial dependence, the residual gauge freedom are simply constant shifts in the gauge field, under which $\mu$ is trivially invariant.

We are interested in expressing our result in terms of the actual magnetic field, the one that we can measure. In the coordinates that we are working, the magnetic field does not correspond exactly with the components of the electromagnetic tensor, contrary to what happens in Cartesian coordinates, but they are related by\footnote{This is the usual definition of the magnetic field in the hydrodynamics context $B^\rho = \frac{1}{2}\epsilon^{\rho\lambda\mu\nu} u_\lambda F_{\mu\nu} $ once we sit in the local rest frame
of the fluid $u_\mu= (1,0,0,0)$.}:

\begin{equation}
\label{eq:magneticfield}
B^i = \frac{1}{2}\frac{\epsilon(i j k)}{\sqrt{-g}}F_{j k} \quad,
\end{equation}

\noindent
where latin indices run over spatial coordinates. This gives the promised radial magnetic field $B(r) \equiv B^r(r) = \frac{Q}{r^2}\, $. 

We emphasize again that the background need not satisfy any equations of motion. As part of the background construction one assumes that there is a horizon $f(r_h) = 0$ and asymptotic flatness:  $f(r\to \infty)=1\,$ and $f'(r\to \infty)=f''(r\to \infty)=0\,$. 
\par 

We can now evaluate the anomaly in the same fashion of the last section. For simplicity of notation we consider a single right-handed fermion. The generalization to a theory with anomaly coefficient $d_{abc}$ is straightforward. We find the following result:

\begin{equation}
\label{eq:NumberCurrentdiv}
    \nabla_\mu J^\mu =\dfrac{1}{\sqrt{-g}}\partial_{\mu}(\sqrt{-g}J^{\mu})=  Q\dfrac{A_t'}{4\pi^2r^2}\,.
\end{equation}

Notice that the Pontryagin density vanishes in this background and only the first piece of the anomaly in (\ref{eq:anomalyJ}) contributes. Once we multiply both sides by $\sqrt{-g}=r^2 \sin\theta$ the result may be integrated over the region between two concentric spheres, one at the horizon and the other one at infinity. Then, we may use Stokes' theorem and imposing that the current vanishes on the horizon gives

\begin{equation}
\left. \left( r^2 J^r \right) \right|_\infty =   \frac{Q}{4 \pi^2} (A_t(\infty) - A_t(r_h)) = \dfrac{Q}{4\pi^2}\mu \,.
\end{equation}

This result is a bit trickier than the one obtained in section \ref{sec:2.1}. 
If we take a look at the l.h.s. of the equation we may notice that the factor $r^2$ coming from $\sqrt{-g}$ diverges at infinity. We may momentarily introduce a cutoff scale $\Lambda$ up to which we integrate in the radial direction to explore the behavior of the current at infinity. Again we impose the boundary condition that the current at the horizon vanishes. Thus we find 

\begin{equation}
\label{eq:CMELambda}
    J^r(\Lambda) =   \frac{1 }{4 \pi^2} \mu(\Lambda) B^r(\Lambda) \quad .
\end{equation}
Where $\mu(\Lambda) = A_t(\Lambda) - A_t(r_h)$. We may notice that this magnetic field decays at infinity, so when we take $\Lambda \to \infty$ the whole term goes to zero. The decay of the field at infinity is indeed a reasonable physical condition. Nonetheless, this does not mean that there is no CME current. To solve the puzzle let us integrate the current over a spherical shell and take the radius to infinity, that is the flux of the current through a spherical surface. The result is found to be

\begin{equation}
    \Phi= \Lambda^2 \int J^r d\Omega_{(2)}^2  = \frac{Q}{\pi} \mu \quad ,
\end{equation}

\noindent which is obviously non-vanishing. One is thus led to ignore the fact that the magnetic field vanishes at infinity and write the chiral magnetic effect in terms of $B_{\infty}$, which is to be understood as $B(\Lambda)$ in equation (\ref{eq:CMELambda}). With this, we can conclude that the current obtained from the integration of the anomaly

\begin{equation}
\label{eq:CMEJ}
    \vec{j}_{CME} = \vec{J}(\infty) =  \pm \frac{\mu_{R,\:\:L}}{4\pi^2} \vec{B}_{\infty} \quad ,
\end{equation}

\noindent describes correctly the CME of a chiral $U(1)$ current.

Let us now consider the energy current $J_E^\mu = T^{\mu\nu} \eta_\nu$, with $\eta = \partial_t$ the timelike Killing vector. 
Symmetry of the energy-momentum tensor and the Killing property of $\nabla_{(\mu} \eta_{\nu)}=0$ imply

\begin{equation}
\label{eq:divJE}
    (\nabla_{\mu} T^{\mu \nu})\eta_{\nu} = \nabla_{\mu}( T^{\mu \nu}\eta_\nu) - T^{\mu\nu}\nabla_{\mu}\eta_{\nu}= \nabla_{\mu} J_E^\mu\,.
\end{equation}

\noindent
In order to explicitly evaluate equation (\ref{eq:anomalyT}) in the background (\ref{eq:CMEbackground}) we first need the current as a function of $r$, not only the value at infinity that we extracted previously. Integrating equation (\ref{eq:NumberCurrentdiv}) up to a generic point $r$ it is immediate to find $4\pi^2 J^r = Q (A_t(r)-A_t|_{r_h})/r^2 $.  Including the later current into (\ref{eq:anomalyT}) and further contracting with the killing vector gives

\begin{equation}
    \nabla_\mu J_E^{\mu}  = \dfrac{1}{\sqrt{-g}}\partial_{\mu}(\sqrt{-g}J_E^\mu)= Q\dfrac{(A_t-A_t|_{r_h})A_t'}{4\pi^2r^2} + Q\dfrac{f f'''}{192\pi^2r^2}\,.
\end{equation}

The first contribution arises from the classical part of the energy-momentum tensor divergence, whereas the second term stems from the anomalous part. Repeating the same integration proccess as before, now from the horizon up to a generic point $r$ and imposing again that the current vanishes at the horizon we find

\begin{equation}
    \left.\left( r^2 J^r_E \right) \right|_r= Q\dfrac{(A_t(r)-A_t(r_h))^2 }{8\pi^2} + Q\dfrac{f(r)f''(r)-\frac{1}{2}f'(r)^2}{192\pi^2} +Q\dfrac{f'(r_h)^2}{348\pi^2}\,.
\end{equation}

Evaluating at infinity and using (i) the definition of the chemical potential (ii) the Hawking temperature and (iii) the same trick as before to bypass the vanishing magnetic field, we are left with

\begin{equation} 
\label{eq:CMET}
\vec{j}_{E, \: CME} = \vec{J}_E(\infty) = \pm \left(\frac{\mu_{R,\:L}^2}{8\pi^2} + \frac{T^2}{24}\right) \vec{B}_{\infty} \,.
\end{equation}
Also in this case the expression coincides with the relevant term in (\ref{eq:cmecveT}).

Which contains contributions from both the gauge and the gravitational anomaly as expected.

Notice again that even though the magnetic field decays to $0$ at infinity, the flux across a spherical shell at infinity

\begin{equation} \label{3-Energy-Flux}
\Phi_E = \int_0^\pi d\theta \int_0^{2\pi} d\phi \Lambda^2 \sin \theta J_E^r = \pm 4 \pi Q \left(\frac{\mu_{R,\:L}^2}{8\pi^2} + \frac{T^2}{24}\right) 
\end{equation}

\noindent remains finite.

As a comment we would like to point out that both (\ref{eq:CMEJ}) and (\ref{eq:CMET}) are also reproduced by choosing a background gauge field with a uniform magnetic field pointing in the $z$-th direction (strong magnetic field limit). The metric background, though generic, should be consistent with the previous choice, and this would amount to having a metric that breaks the $SO(3)$ rotational symmetry down to $SO(2)$. A good candidate for such a metric is the $\Omega \to 0$ limit of equation (\ref{eq:CVEbackground}) with possibly a generic function $h(z)$ with the right asymptotics respecting the isometries on the transverse plane:   

\begin{equation}
ds^2 = -f(z) dt^2 + \frac{dz^2}{f(z)} + h(z)(d\rho^2 + \rho^2 d\phi^2) \,.
\end{equation}

It has been written in cylindrical coordinates for convenience. The function $h(z)$ however does not modify the results provided that we normalize $h(z\to \infty) = 1$. 
The fact that both a radial decaying magnetic field or an axial constant magnetic field give the right result may be a consequence of the universality of the calculation. However, it is not clear whether one will be preferred over the other if the same method is applied to higher spin currents, for which the procedure is analogous to the number current and the energy momentum tensor.

\section{Chiral Vortical Effect in 3+1 dimensions via Anomaly}

We extend now the analysis to include the contribution of a $U(1)$ chemical potential in the chiral vortical effect. Though the procedure is by now clear, the choice of a suitable background plays a central role. For the background metric we still use (\ref{eq:CVEbackground}):

\begin{equation}
\label{eq:CVEbackground2}
ds^2 = -f(z) \frac{(dt - \Omega \rho^2 d\phi)^2}{1-\Omega^2\rho^2} + \frac{dz^2}{f(z)} + d\rho^2 + \frac{ \rho^2(d\phi - \Omega dt)^2}{1-\Omega^2\rho^2} \,.
\end{equation} 

The background comes along with the asymptotics for the generic function. Repeating previous arguments we assume the existence of a horizon at some point $z_h$ as well as asymptotic flatness:$f(z\to \infty)=1\,$ and $f'(z\to \infty)=f''(z\to \infty)=0\,$. As for the gauge field recall that the background, though generic, should be able to solve the equations of motion if so required. Taking a look at the field strength of the Kerr-Newmann solution we may infer that a good candidate would be 

\begin{equation}
	A = A_t(z)dt + \Omega A_{\phi}(\rho,z)d\phi\,, 
\end{equation}

\noindent
where we are again working in the axial gauge $A_z=0$ and, as explained above, restricting to stationary solutions. Thus, the residual gauge freedom for the time component of the gauge field reduces to constant shifts. However, it is not hard to convince oneself that the ansatz is too generic. By looking at the examples in the previous sections, the only functions that are kept generic encode the information about the state: temperature and chemical potential; yet the $\phi$ component does not contain such information. Actually, it represents the magnetic field induced by charged particles rotating around the $z$-th axis. We may use this notion to factorize the $\rho$ dependence as $A_{\phi}=-\rho^2 a_{\phi}(z)$. Again $a_{\phi}$ is too generic. Taking a look at Maxwell's equations of motion suggests that $a_{\phi}(z)=A_t(z)-G$, with $G$ some integration constant. In particular,

\begin{equation}
    \nabla_{\mu}F^{\mu t} = 0 
\end{equation}
\begin{equation}
    \nabla_{\mu}F^{\mu \phi} = \Omega \nabla_{\mu}F^{\mu t} + \Omega \left(f(A_t'-a_{\phi}')\right)' = 0\,.
\end{equation}

\noindent
Consequently $f(A_t'-a_{\phi}')=const$, and evaluating it at the horizon gives $const=0$. A final integration gives the sought result $a_{\phi}(z)=A_t(z)-G$. We will later rederive the same result without making use of the equations of motion.
 
 Notice that a constant shift in $A_t$ may be reabsorbed in a constant shift in $G$, so that $a_{\phi}$ respects the residual gauge invariance. Consequently, we rewrite the generic background as

\begin{equation}
\label{eq:CVEgauge}
	A = (A_t(z)-G)(dt - \rho^2\Omega d\phi)\,, 
\end{equation}

\noindent
where the constant $G$ has been added also to the temporal component without loss of generality to make manifest that the one-form is proportional to the combination that appears in the metric (\ref{eq:CVEbackground2}) and hence respects the symmetry associated to the gravitomagnetic gauge field. In fact, this form of the gauge field
is easiest to understand by noting that there is a local velocity field $v_\phi = \rho^2 d\phi$ due to frame dragging. The rotation is then introduced by $dt \rightarrow dt-\rho^2 \Omega d\phi$ and $d\phi \rightarrow d\phi - \Omega dt$. 
 An alternative derivation of (\ref{eq:CVEgauge}) consists in requiring that the background respects the gravitomagnetic gauge symmetry $t \rightarrow t +\lambda$ and $A_g \rightarrow A_g + d\lambda$ where $\lambda$ is a $t$ independent gauge parameter. This suggests that only the the differential one form $dt- A_g$ appears.
 Then one notices that rotation to lowest order in $\Omega$ is represented by
 the gravitomagnetic field $A_g = \rho^2 \Omega d\phi$. Since we are interested 
 only in the terms linear in $\Omega$ this is sufficient for our purposes.
 
 The magnetic field in the $z$-th direction is obtained using (\ref{eq:magneticfield}):  $B^z(z)=2\Omega (G-A_t(z))$. As boundary conditions for the computation we shall demand that the magnetic field vanishes at infinity. This condition means that there is no additional magnetic field at infinity. We insist that it is imposed in the combination $A_t(z)-G$ and hence respects the residual gauge invariance. Solving the condition for the magnetic field reveals $G=A_t|_{\infty}\,$.

Now that the background has been constructed we may proceed to integrate the anomaly equation. Following the steps of \cite{Stone:2018zel} presented in section \ref{sec:2.2} we shall assume that only the $z$-th component of the current is needed. In the discussion section we shall comment on the limitations of this assumption. Besides, the computation shall be done on-axis and to leading order in $\Omega$, for higher contributions are not universal as explained above. Substituting the ansatz (\ref{eq:CVEbackground2}) and (\ref{eq:CVEgauge}) into the anomaly equation (\ref{eq:anomalyJ}) specified to a single right-handed $U(1)$ current we find

\begin{equation}
	\nabla_\mu J^{\mu} = \dfrac{1}{\sqrt{-g}}\partial_{\mu}(\sqrt{-g}J^{\mu}) =\partial_z J^z =  -\frac{  (A_t(z)-A_t|_{\infty}) A_t'(z)}{2 \pi
   ^2}\Omega-\frac{  f'(z) f''(z)}{96 \pi ^2}\Omega + \mathcal{O}(\Omega^3)\,.
\end{equation}

Integrating the previous equation gives

\begin{equation}
\label{eq:CVEcurrentz}
	J^z(z)  = -\dfrac{(A_t(z)-A_t|_{\infty})^2}{4 \pi^2}\Omega - \dfrac{f'(z)^2}{192\pi^2}\Omega + \left(\dfrac{\mu^2}{4 \pi^2}+\dfrac{T^2}{12}\right)\Omega + \mathcal{O}(\Omega^3)\,,
\end{equation}

\noindent
where we have written the contribution from the horizon as a function of chemical potential (\ref{eq:chemical}) and the Hawking temperature. Evaluating at infinity and keeping the leading piece in $\Omega$ we obtain the CVE in presence of temperature and chemical potential:

\begin{equation}\label{eq:cvej}
	j_{CVE}=J^z(\infty) = \left(\dfrac{\mu^2}{4 \pi^2}+\dfrac{T^2}{12}\right)\Omega \,.
\end{equation}

We now proceed with the computation for the energy current which, as before, is defined  out of the energy-momentum tensor contracted with a timelike Killing vector (\ref{eq:Energycurrent}). Evaluating the anomalous conservation equation (\ref{eq:anomalyT}) in the rotating background (\ref{eq:CVEbackground2}, \ref{eq:CVEgauge}) and using equations (\ref{eq:divJE}) and (\ref{eq:CVEcurrentz}) we find

\begin{equation}
\begin{split}
	\nabla_{\mu}J^{\mu}_E &= \dfrac{1}{\sqrt{-g}}\partial_\mu(\sqrt{-g}J^\mu_E) = \partial_z J_E^z = \left(\frac{1}{12} T^2   A_t' + \mu ^2\frac{   A_t'}{4 \pi ^2}-\frac{ 
   (A_t-A_t|_{\infty})^2 A_t'}{4 \pi ^2}\right)\Omega  \\& -\dfrac{1}{48\pi^2}\left(\frac{ 1}{2}A_t'' f  f'+ A_t'\left[ f 
   f''+\frac{1}{4}  f'\,^2\right]+\frac{1}{2}(A_t-A_t|_\infty) f
   f'''\right)\Omega + \mathcal{O}(\Omega^3)\,.
\end{split}
\end{equation}

As it is expected, the r.h.s. gives a total derivative and the result only depends on the boundary conditions of the background functions. Following the prescription given above, we integrate: 

\begin{equation}
\begin{split}
\label{eq:CVEcurrentenergy}
    J_E^z(z) & = \left(\frac{1}{12}
   T^2   \overline{A}_t+\frac{\mu ^2   }{4 \pi ^2}\overline{A}_t-\frac{
   \overline{A}_t^3}{12 \pi ^2}\right)\Omega
   \\ & -\left(  \overline{A}'_tf f'+\overline{A}_t\left[f f''-\frac{
   1}{2}f'\,^2\right]\right)\dfrac{1}{96\pi^2} \Omega + \mathcal{O}(\Omega^3) \,,
\end{split}
\end{equation}

\noindent
where we have defined $\overline{A}_t = A_t(z)-A_t|_{\infty}$ for the sake of visual simplicity. Under the same assumptions presented before we recover the CVE for the energy current:

\begin{equation}
     j_{E,\: CVE} = J_E^z(\infty) =  \left(\dfrac{1}{6}\mu T^2 + \dfrac{1}{6\pi^2}\mu^3\right)\Omega\,.
\end{equation}

Collecting our results for the chiral vortical effects we find

\begin{equation}
\begin{split}
  \vec{j}_{CVE} = &  \left(\dfrac{\mu^2}{4 \pi^2}+\dfrac{T^2}{12}\right) \vec{\Omega} \,, \\
  \vec{j}_{E,\: CVE} = & \left(\dfrac{1}{6}\mu T^2 + \dfrac{1}{6\pi^2}\mu^3\right) \vec{\Omega}\,,
\end{split}
\end{equation}
which coincides with the relevant terms in equations (\ref{eq:cmecveJ}) and (\ref{eq:cmecveT}).
\section{Summary and Discussion}
We have revisited the method of integrating the anomaly equation in black hole backgrounds to derive anomaly induced currents at finite temperature and chemical potential. The new results include the derivation for weak magnetic fields and 
for the chiral vortical effect in the energy current.
The derivations for gauge and gravitational anomalies proceed exactly in the same manner
and can be integrated straightforwardly. One obtains the well-known coefficients of
chiral magnetic and chiral vortical effects once vanishing of the current at the horizon
is imposed as a boundary condition.

 We turn our attention now to the assumption the CVE current is parallel to the $z$-th direction. Interestingly, this is not necessarily the case. If we allow in equation (\ref{eq:CVEcurrentz}) for a generic current depending on $z$ and $\rho$\footnote{Dependence on time or on the azimuthal angle breaks the symmetries of the problem.} then a new contribution may arise from the $\rho$ component of the current:\footnote{$J^z$ may also depend on the radial component $\rho$. Since we are interested in the current on the rotation axis we could write $J^z(z,\rho)\simeq J^z(z,0) + \mathcal{O}(\rho)$, where the higher order terms do not survive the limit $\rho \to 0$.}

\begin{equation}
\label{eq:CVEcurrentz2}
\nabla_\mu J^{\mu} = \partial_z J^z(z) + \dfrac{1}{\rho}\partial_{\rho}\left(\rho J^{\rho}(\rho,z)\right) =  -\frac{  (A_t(z)-A_t|_{\infty}) A_t'(z)}{2 \pi
   ^2}\Omega-\frac{  f'(z) f''(z)}{96 \pi ^2}\Omega + \mathcal{O}(\Omega^3)\,.
\end{equation}

\noindent
Clearly $J^{\rho}$ gives a finite contribution so long as $J^{\rho}=\rho g(z)+\mathcal{O}(\rho^2)$, for an arbitrary function $g(z)$. Thus, the anomaly equation could be integrated with arbitrary contributions from $J^{\rho}$ and $J^z$ which balance each other. This freedom is not surprising from a mathematical point of view, since the divergence of a vector field does not uniquely determine the vector field. One needs to know also the curl. In two dimensions or in the CME derivation in section \ref{sec:cme}, the amount of symmetry was enough to univoquely integrate the anomaly equation, but for the CVE in $3+1$ dimensions this is no longer true. Therefore, one needs to include some ad-hoc knowledge (namely, the current is parallel to the $z$-th direction) to recover the well-known CVE result. Similar considerations apply for the CVE energy current.
In this respect we also want to comment on the relation between the chiral magnetic effect in the energy current eq. (\ref{eq:CMET}) and the chiral vorical effect for the current eq. (\ref{eq:cvej}).
It follows from basic principles of quantum field theory that these two coefficients are the same. As pointed out some time ago in \cite{Amado:2011zx} these transport coefficients can be computed using Kubo formulas in which only roles of the energy and U(1) currents are exchanged. In the particular kinematic limit at zero frequency these have to be the same by complex conjugation.
It should be noted that in our calculations the chiral magnetic and chiral vortical coefficients arise from integration in quite different geometries. Using however the general fact that the two coefficients have to be the same this does give another argument why the $\rho$ component of the current in (\ref{eq:CVEcurrentz2}) has to vanish once the chiral magnetic effect in the energy current (\ref{eq:CMET}) is known.

The method seems to be universally applicable for all theories that feature chiral anomalies. It is however well-known that the method fails for spin $3/2$ fields \cite{Loganayagam:2012zg}. 
In this respect it is important to note that the axial current of a relativistic massless
spin $3/2$ field is not invariant under the fermionic gauge symmetry and therefore not
a good local observable. Taking this into account it is probably not too surprising that the gravitational anomaly equation does not reproduce the CVE in that case. On the other hand a new theory for a relativistic spin $3/2$ field with a set of auxiliary fields was introduced in \cite{Adler:2015yha}. This theory allows a perturbative coupling to gauge fields and 
has a gauge invariant axial current. Unfortunately the quantum theory still suffers from 
pathologies such as a non-definite Hilbert space \cite{Adler:2017lki}. Nevertheless its anomalous transport properties have been recently studied in a series of papers \cite{Prokhorov:2021bbv} and the relation to the anomaly integration method has been studied \cite{Prokhorov:2022rna}. 

Similarly there is a gravitational anomaly in spin one gauge fields \cite{Vainshtein:1988ww,Agullo:2016lkj} but again the corresponding current (the helicity current) is not a gauge invariant local operator. Interestingly in that case it has been shown that the integration of the anomaly equation does reproduce the CVE for
the helicity current if a mass as regulator of infrared divergencies is introduced \cite{Prokhorov:2020npf}.  In this respect it would be interesting to study if the 
anomaly integration method will also fix chiral magnetic and chiral vortical effects
in the higher bosonic and fermionic Zilch currents \cite{Chernodub:2018era,Copetti:2018mxw,Alexandrov:2020zsj}

It should also be noted that a global gravitational anomaly has
been invoked to fix the temperature dependence of the CVE coefficient in \cite{Golkar:2015oxw, Glorioso:2017lcn}. Unfortunately it is so far not known how this argument can be generalized to the
spin one case or to the enlarged spin $3/2$ theory. 

We also note that upon finishing this manuscript \cite{Frolov:2022rod} appeared which also studies integration of the anomaly equation in black hole geometries.

\section*{Acknowledgements}
This work has been supported bySpanish National Plan projects
PGC2018-095976-B-C21 MCIN, AEI, FEDER, 
PGC2018-095862-B-C21 MCIN, AEI, FEDER,
Severo Ochoa Center of Excellence CEX2020-001007-S
. The work of PSB is supported by Fundaci\'on S\'eneca, Agencia de Ciencia y Tecnolog\'ia de la Regi\'on de Murcia, grant 21609/FPI/21 and Spanish Ministerio de Ciencia e Innovación PID2021-125700NA-C22. SMT is supported by a FPI-UAM predoctoral fellowship.

\bibliography{AnomCur}{}

\begin{thebibliography}{10}

\bibitem{Bertlmann:1996xk}
R.~A. Bertlmann.
\newblock {\em {Anomalies in quantum field theory}}.
\newblock 1996.

\bibitem{Landsteiner:2016led}
Karl Landsteiner.
\newblock {Notes on Anomaly Induced Transport}.
\newblock {\em Acta Phys. Polon. B}, 47:2617, 2016.

\bibitem{Kharzeev:2013ffa}
Dmitri~E. Kharzeev.
\newblock {The Chiral Magnetic Effect and Anomaly-Induced Transport}.
\newblock {\em Prog. Part. Nucl. Phys.}, 75:133--151, 2014.

\bibitem{Bardeen:1984pm}
William~A. Bardeen and Bruno Zumino.
\newblock {Consistent and Covariant Anomalies in Gauge and Gravitational
  Theories}.
\newblock {\em Nucl. Phys. B}, 244:421--453, 1984.

\bibitem{Vilenkin:1980fu}
A.~Vilenkin.
\newblock {EQUILIBRIUM PARITY VIOLATING CURRENT IN A MAGNETIC FIELD}.
\newblock {\em Phys. Rev. D}, 22:3080--3084, 1980.

\bibitem{Vilenkin:1979ui}
A.~Vilenkin.
\newblock {MACROSCOPIC PARITY VIOLATING EFFECTS: NEUTRINO FLUXES FROM ROTATING
  BLACK HOLES AND IN ROTATING THERMAL RADIATION}.
\newblock {\em Phys. Rev. D}, 20:1807--1812, 1979.

\bibitem{Giovannini:1997eg}
Massimo Giovannini and M.~E. Shaposhnikov.
\newblock {Primordial hypermagnetic fields and triangle anomaly}.
\newblock {\em Phys. Rev. D}, 57:2186--2206, 1998.

\bibitem{Alekseev:1998ds}
Anton~Yu. Alekseev, Vadim~V. Cheianov, and Jurg Frohlich.
\newblock {Universality of transport properties in equilibrium, Goldstone
  theorem and chiral anomaly}.
\newblock {\em Phys. Rev. Lett.}, 81:3503--3506, 1998.

\bibitem{Fukushima:2008xe}
Kenji Fukushima, Dmitri~E. Kharzeev, and Harmen~J. Warringa.
\newblock {The Chiral Magnetic Effect}.
\newblock {\em Phys. Rev. D}, 78:074033, 2008.

\bibitem{Son:2009tf}
Dam~T. Son and Piotr Surowka.
\newblock {Hydrodynamics with Triangle Anomalies}.
\newblock {\em Phys. Rev. Lett.}, 103:191601, 2009.

\bibitem{Neiman:2010zi}
Yasha Neiman and Yaron Oz.
\newblock {Relativistic Hydrodynamics with General Anomalous Charges}.
\newblock {\em JHEP}, 03:023, 2011.

\bibitem{Prokhorov:2022udo}
G.~Yu. Prokhorov, O.~V. Teryaev, and V.~I. Zakharov.
\newblock {Hydrodynamic Manifestations of Gravitational Chiral Anomaly}.
\newblock {\em Phys. Rev. Lett.}, 129(15):151601, 2022.

\bibitem{Prokhorov:2022snx}
Georgy~Yu. Prokhorov, Oleg~V. Teryaev, and Valentin~I. Zakharov.
\newblock {Gravitational chiral anomaly and anomalous transport for fields with
  spin 3/2}.
\newblock {\em Phys. Lett. B}, 840:137839, 2023.

\bibitem{Kimura:2011ef}
Taro Kimura and Tatsuma Nishioka.
\newblock {The Chiral Heat Effect}.
\newblock {\em Prog. Theor. Phys.}, 127:1009--1017, 2012.

\bibitem{Landsteiner:2011cp}
Karl Landsteiner, Eugenio Megias, and Francisco Pena-Benitez.
\newblock {Gravitational Anomaly and Transport}.
\newblock {\em Phys. Rev. Lett.}, 107:021601, 2011.

\bibitem{Landsteiner:2011iq}
Karl Landsteiner, Eugenio Megias, Luis Melgar, and Francisco Pena-Benitez.
\newblock {Holographic Gravitational Anomaly and Chiral Vortical Effect}.
\newblock {\em JHEP}, 09:121, 2011.

\bibitem{Erdmenger:2008rm}
Johanna Erdmenger, Michael Haack, Matthias Kaminski, and Amos Yarom.
\newblock {Fluid dynamics of R-charged black holes}.
\newblock {\em JHEP}, 01:055, 2009.

\bibitem{Banerjee:2008th}
Nabamita Banerjee, Jyotirmoy Bhattacharya, Sayantani Bhattacharyya, Suvankar
  Dutta, R.~Loganayagam, and P.~Surowka.
\newblock {Hydrodynamics from charged black branes}.
\newblock {\em JHEP}, 01:094, 2011.

\bibitem{Jensen:2012kj}
Kristan Jensen, R.~Loganayagam, and Amos Yarom.
\newblock {Thermodynamics, gravitational anomalies and cones}.
\newblock {\em JHEP}, 02:088, 2013.

\bibitem{Stone:2018zel}
Michael Stone and Jiyoung Kim.
\newblock {Mixed Anomalies: Chiral Vortical Effect and the Sommerfeld
  Expansion}.
\newblock {\em Phys. Rev. D}, 98(2):025012, 2018.

\bibitem{Robinson:2005pd}
Sean~P. Robinson and Frank Wilczek.
\newblock {A Relationship between Hawking radiation and gravitational
  anomalies}.
\newblock {\em Phys. Rev. Lett.}, 95:011303, 2005.

\bibitem{Jensen:2013rga}
Kristan Jensen, R.~Loganayagam, and Amos Yarom.
\newblock {Chern-Simons terms from thermal circles and anomalies}.
\newblock {\em JHEP}, 05:110, 2014.

\bibitem{Murata:2006pt}
Keiju Murata and Jiro Soda.
\newblock {Hawking radiation from rotating black holes and gravitational
  anomalies}.
\newblock {\em Phys. Rev. D}, 74:044018, 2006.

\bibitem{Landsteiner:2011tg}
Karl Landsteiner, Eugenio Megias, and Francisco Pena-Benitez.
\newblock {Anomalies and Transport Coefficients: The Chiral Gravito-Magnetic
  Effect}.
\newblock In {\em {11th Workshop on Non-Perturbative Quantum Chromodynamics}},
  10 2011.

\bibitem{Hawking:1995ap}
S.~W. Hawking and Simon~F. Ross.
\newblock {Duality between electric and magnetic black holes}.
\newblock {\em Phys. Rev. D}, 52:5865--5876, 1995.

\bibitem{Amado:2011zx}
Irene Amado, Karl Landsteiner, and Francisco Pena-Benitez.
\newblock {Anomalous transport coefficients from Kubo formulas in Holography}.
\newblock {\em JHEP}, 05:081, 2011.

\bibitem{Loganayagam:2012zg}
R.~Loganayagam.
\newblock {Anomalies and the Helicity of the Thermal State}.
\newblock {\em JHEP}, 11:205, 2013.

\bibitem{Adler:2015yha}
Stephen~L. Adler.
\newblock {Classical Gauged Massless Rarita-Schwinger Fields}.
\newblock {\em Phys. Rev. D}, 92(8):085022, 2015.

\bibitem{Adler:2017lki}
Stephen~L. Adler, Marc Henneaux, and Pablo Pais.
\newblock {Canonical Field Anticommutators in the Extended Gauged
  Rarita-Schwinger Theory}.
\newblock {\em Phys. Rev. D}, 96(8):085005, 2017.

\bibitem{Prokhorov:2021bbv}
G.~Yu. Prokhorov, O.~V. Teryaev, and V.~I. Zakharov.
\newblock {Chiral vortical effect in extended Rarita-Schwinger field theory and
  chiral anomaly}.
\newblock {\em Phys. Rev. D}, 105(4):L041701, 2022.

\bibitem{Prokhorov:2022rna}
G.~Yu. Prokhorov, O.~V. Teryaev, and V.~I. Zakharov.
\newblock {Gravitational chiral anomaly for spin 3/2 field interacting with
  spin 1/2 field}.
\newblock {\em Phys. Rev. D}, 106(2):025022, 2022.

\bibitem{Vainshtein:1988ww}
A.~I. Vainshtein, A.~D. Dolgov, Valentin~I. Zakharov, and I.~B. Khriplovich.
\newblock {CHIRAL PHOTON CURRENT AND ITS ANOMALY IN A GRAVITATIONAL FIELD}.
\newblock {\em Sov. Phys. JETP}, 67:1326--1332, 1988.

\bibitem{Agullo:2016lkj}
I.~Agullo, A.~del Rio, and J.~Navarro-Salas.
\newblock {Electromagnetic duality anomaly in curved spacetimes}.
\newblock {\em Phys. Rev. Lett.}, 118(11):111301.

\bibitem{Prokhorov:2020npf}
G.~Yu Prokhorov, O.~V. Teryaev, and V.~I. Zakharov.
\newblock {Chiral vortical effect for vector fields}.
\newblock {\em Phys. Rev. D}, 103(8):085003, 2021.

\bibitem{Chernodub:2018era}
M.~N. Chernodub, Alberto Cortijo, and Karl Landsteiner.
\newblock {Zilch vortical effect}.
\newblock {\em Phys. Rev. D}, 98(6):065016, 2018.

\bibitem{Copetti:2018mxw}
Christian Copetti and Jorge Fern\'andez-Pend\'as.
\newblock {Higher spin vortical Zilches from Kubo formulae}.
\newblock {\em Phys. Rev. D}, 98(10):105008, 2018.

\bibitem{Alexandrov:2020zsj}
Artem Alexandrov and Pavel Mitkin.
\newblock {Zilch Vortical Effect for Fermions}.
\newblock {\em JHEP}, 05:070, 2021.

\bibitem{Golkar:2015oxw}
Siavash Golkar and Savdeep Sethi.
\newblock {Global Anomalies and Effective Field Theory}.
\newblock {\em JHEP}, 05:105, 2016.

\bibitem{Glorioso:2017lcn}
Paolo Glorioso, Hong Liu, and Srivatsan Rajagopal.
\newblock {Global Anomalies, Discrete Symmetries, and Hydrodynamic Effective
  Actions}.
\newblock {\em JHEP}, 01:043, 2019.

\bibitem{Frolov:2022rod}
Valeri~P. Frolov, Alex Koek, Jose~Pinedo Soto, and Andrei Zelnikov.
\newblock {Chiral anomalies in black hole spacetimes}.
\newblock 12 2022.

\end{thebibliography}
\bibliographystyle{unsrt}

\end{document}